\documentclass[layout=traditional]{achemso}
\usepackage{amssymb,amsmath}
\usepackage{graphicx} 

\usepackage[unicode,colorlinks=true,linkcolor=blue,citecolor=blue,urlcolor=blue]{hyperref}
\usepackage[all]{hypcap} 

\newcommand{\kB}{k_{\rm\scriptscriptstyle B}} 

\author{David Vor{\' a}{\v c}}
\affiliation{Charles University, 
Faculty of Mathematics and Physics, 
Department of Macromolecular Physics, 
V Hole\v{s}ovi\v{c}k\'ach 2, 
CZ-18000 Praha 8, 
Czech Republic}

\author{Philipp Maass}
\email{philipp.maass@uos.de}
\affiliation{ 
Universit{\" a}t Osnabr{\" u}ck, 
Fachbereich Physik, 
Barbarastra{\ss}e 7, 
D-49076 Osnabr{\" u}ck, 
Germany}

\author{Artem Ryabov}
\email{rjabov.a@gmail.com}
\affiliation{Charles University, 
Faculty of Mathematics and Physics, 
Department of Macromolecular Physics, 
V Hole\v{s}ovi\v{c}k\'ach 2, 
CZ-18000 Praha 8, 
Czech Republic}

\title{Cycle completion times probe interactions\\ with environment}

\SectionNumbersOn 
\begin{document}

\begin{abstract}
Recent measurements of durations of non-equilibrium processes provide valuable information on microscopic mechanisms and energetics. Comprehensive theory for corresponding experiments so far is well developed for single-particle systems only. Little is known for interacting systems in non-equilibrium environments. Here, we introduce and study a basic model for cycle processes interacting with an environment that can
exhibit a net particle flow. We find a surprising richness of cycle time variations with environmental conditions. This manifests itself in
unequal cycle times $\tau^+$ and $\tau^-$  in forward and backward cycle direction with both 
asymmetries $\tau^-<\tau^+$ and $\tau^->\tau^+$, speeding up of backward cycles
by interactions, and dynamical phase transitions, where cycle times become multimodal functions of the
bias. The model allows us to relate these effects to specific
microscopic mechanisms, which can be helpful for interpreting experiments.
\end{abstract} 
\maketitle
\newpage 


With the rapid advancement of experimental techniques,
it has become possible to analyze individual
molecular processes with unprecedented precision. 
Corresponding investigations test fundamental assumptions of theories 
and provide new insights into microscopic mechanisms of 
the kinetics of activated processes.
For example, the long-standing Kramers' theory of barrier crossing could be tested
by following the folding and unfolding of single biological molecules \cite{Neupane/etal:2018JPCB}, like proteins \cite{Chung:2013} 
and nucleic acids \cite{Neupane/etal:2016, Neupane/etal:2017}. Detailed information could be obtained for the
transition paths, e.g.\
local velocities  \cite{Neupane/etal:2018} and
average shapes \cite{Hoffer/etal:2019}.

In such experiments, times for completing activated processes 
in forward and backward  direction, e.g., folding and unfolding,
are random variables. It was shown theoretically that their distributions
are identical,\cite{BerezhkovskiiPRL2006} if 
time-reversibility of the individual trajectories is assumed.
This symmetry of distributions was
demonstrated explicitly 
in a broad class of kinetic models.\cite{AstumianPCCP2007, Lindner2001,  KolomeiskyPRE2005, LINDEN2007, Tsygankov2007,  ZhangJCP2007, SBerezhkovskiiJCP2006, QianPRE2006, BerezhkovskiiJPhysCM2007, DagdugJCP2009, Laleman2017, RednerKrapivksy2018}
Common to all these models is that they refer to the dynamics of a single particle.  
In experimental tests it was observed that the symmetry in single-molecule processes can be broken by a time-dependent external forcing. \cite{Gladrow/etal:2019} 

Recently, it was discovered that the symmetry breaking can be caused also by interactions in many-body systems \cite{Ryabov/etal:2019, Shin/etal:2020}. It has been reported so far for two models, where
thermally activated transitions take place in a constant flow of identical particles with steric exclusion (hard-core) interactions. 
A surprising effect occurs, namely that mean transition times against the flow are always shorter than in flow direction. This raises the question, whether this is a universal feature of transition processes in a nonequilibrium crowded environment. 

Here, instead of elementary transitions, we consider completion times for cyclic processes, as they are relevant in a broad class of situations of biochemistry\cite{Schnakenberg:1976, Hill:1989,Bressloff:2014}, for example, 
in catalytic reactions,\cite{King/Altman:1956,Ge:2008} protein dynamics,\cite{Shukla/etal:2015}
membrane transport,\cite{Hille/Schwarz:1978, Beattie/etal:2018} 
and operation of molecular motors.\cite{Bustamante/etal:2001, Baroncini/etal:2020, Hwang/Hyeon:2017,Hwang/Hyeon:2018} 
These times will be referred to as cycle times in the following. Interactions with the environment
can be both attractive and repulsive, depending on chemical conditions like temperature, salt concentration, pH-value, etc. Moreover, the flow of particles in the environment can be decoupled from the preferred direction of the cycle process. 

\begin{figure}[b!]
\centering
\includegraphics[scale=1]{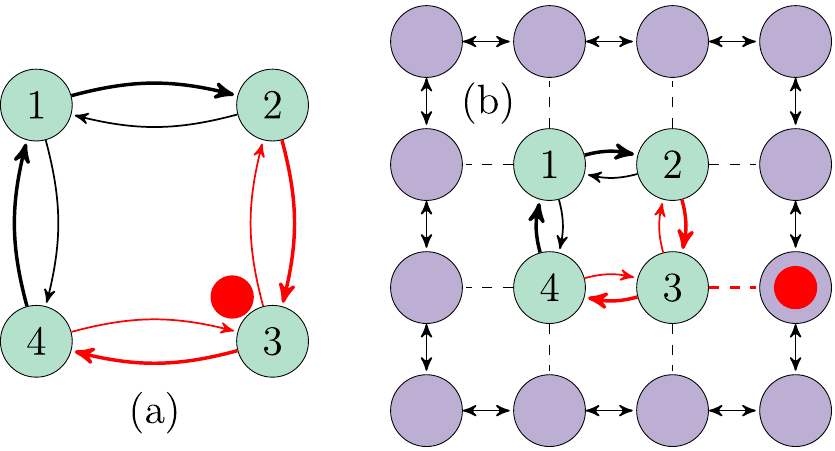} 
\caption{Cyclic chemical processes interacting with an environment: (a) Unicyclic reaction network, where
transition rates are modified by a particle performing a random walk
on the ring. This particle represents the environment's degrees of freedom and is 
at state 3 in the figure (red circle).
(b) Same unicyclic reaction, where the transition rates are modified by diffusing particles in the outer ring.
Dashed lines represent bonds that form 
if a particle in the environment interacts with the system
(one interaction is marked by the red circle). In both (a) and (b),
bold arrows indicate the bias $f>0$ and red arrows modified rates by interactions.}
\label{fig:reservoirmodel}
\end{figure} 


We propose a simple model for such a cyclic process 
which allows us to understand 
how the interplay of system-environment interactions and flow in the surroundings affect cycle times
and the symmetry breaking between cycle times. 
The model  is sketched in Fig.~\ref{fig:reservoirmodel}(a). In fact, it 
is a simplification of the more realistic situation illustrated in Fig.~\ref{fig:reservoirmodel}(b). 

In the model, a reaction coordinate is making transitions between minima of a free energy landscape represented by sites of the inner ring in Fig.~\ref{fig:reservoirmodel}(b). The 
transitions connect different chemical states with free energies $\varepsilon_i$
along the ring. Elementary transition from the state $i$ to a neighboring state $j$ in the cycle occur with rates $k(i\to j )$ which are detailed balanced locally, 
\begin{equation}
\frac{k(i\to j)}{k(j\to i)}=\exp(-\beta\Delta G_{ij}),
\label{eq:dtbalance}
\end{equation}
where $\Delta G_{ij}=\varepsilon_j-\varepsilon_i+f$ is the dissipated free energy in the transition $(i\to j)$ and $1/\beta=\kB T$ is the thermal energy. The constant $f$ is a bias, which can drive the reaction along the cycle in a preferred direction. 

For each actual process, the rate constants
$k(i\to j)$ and $k(j\to i)$ are determined by microscopic mechanisms. Here, we take the widely used exponential form
\begin{equation}
k(i\to j)=a_{ij}\exp\left[-\frac{\beta}{2}(\varepsilon_j-\varepsilon_i+f)\right]\,,
\label{eq:kij}
\end{equation}
where the preexponential factor is symmetric, $a_{ij}=a_{ji}$, and the rate constant $k(j\to i)$ follows from Eq.~\eqref{eq:dtbalance}.

The transition dynamics between individual states can be represented by hopping of a (system) particle. The forward and backward cycle times $\tau^+$ and $\tau^-$ are then determined as follows: 
We initiate a clock when the particle arrives at a certain state, say state 1, and 
record the time when it returns to this state after completing a full cycle in either direction. 
The average time to complete the full cycle in (against) the bias direction is $\tau^+$ ($\tau^-$).
The sampling of cycle times is always done in the steady state.

Interactions with the environment change the state energies. They
can be modeled by environment particles hopping between the states of the outer ring in Fig.~\ref{fig:reservoirmodel}(b) and interacting with the system particle.
We performed extensive kinetic Monte Carlo simulations of this many-body problem.
As it turned out, to capture the full complexity of cycle time behavior it is sufficient to add one ``environment particle'' on the inner ring of states of the system particle.

Such a minimal unicyclic model is illustrated in Fig.~\ref{fig:reservoirmodel}(a).
The environment particle performs a random walk between the cycle states with transition rates as in Eq.~\eqref{eq:kij}. If both particles occupy the same state $i$, the state energy $\varepsilon_i$ is changed by an amount $g$. The rate for the system particle (the reaction coordinate) to transit to a neighboring state $j$ is then given by
\begin{equation}
K({i\to j})=k(i\to j)\exp(\beta g/2)\,,
\end{equation}
that means it is enhanced compared to $k(i\to j)$ for repulsive interactions with the environment particle ($g>0)$ and reduced for attractive interactions ($g<0$). The corresponding backward transitions $K(j\to i)$ follows from the detailed balance condition~\eqref{eq:dtbalance}. 

The transition rates for the environment particle are the same as for the system particle except for the bias $f_{\rm e}$, which instead of $f$ enters the transition rates in Eq.~\eqref{eq:kij}. This bias $f_{\rm e}$ represents a non-equilibrium flow in the environment. The case $f_{\rm e}=f$ corresponds to particle flow in the environment as in the system and for
$g\to+\infty$ resembles the situations in the single file  models studied in previous work, i.e.\ in the BASEP \cite{Ryabov/etal:2019} and ASEP \cite{Shin/etal:2020}. The case $f_{\rm e}=0$ corresponds to an equilibrium environment, where the flow has no preferred direction.
We also consider the case $f_{\rm e}=-f$ representing an anti-flow behavior, which can arise in coupled reaction processes.
Overall, the two-particle model with the three parameters $f$, $f_{\rm e}$ and $g$ can be viewed as a generic minimal model to investigate
influences of system-environment interaction and particle flow in the environment on cycle processes.

We start out by
setting all state energies $\varepsilon_i$ to be equal,
$\varepsilon_i=0$, and also preexponential factors to be the same,  $a_{ij}=a$. 
In all simulations, we set $\beta=1$ and $a=1$, which means that energies are given 
in units of $\kB T$ and times in units of $1/a$. Results were obtained by applying three different methods:
(i) kinetic Monte Carlo simulations, and numerical solutions of (ii) the backward (yielding the mean cycle times)
and (iii) the forward master equations (yielding also distributions of the cycle times).
Details of these procedures can be found in the supporting information.


\begin{figure}[t] 
\includegraphics[scale=1]{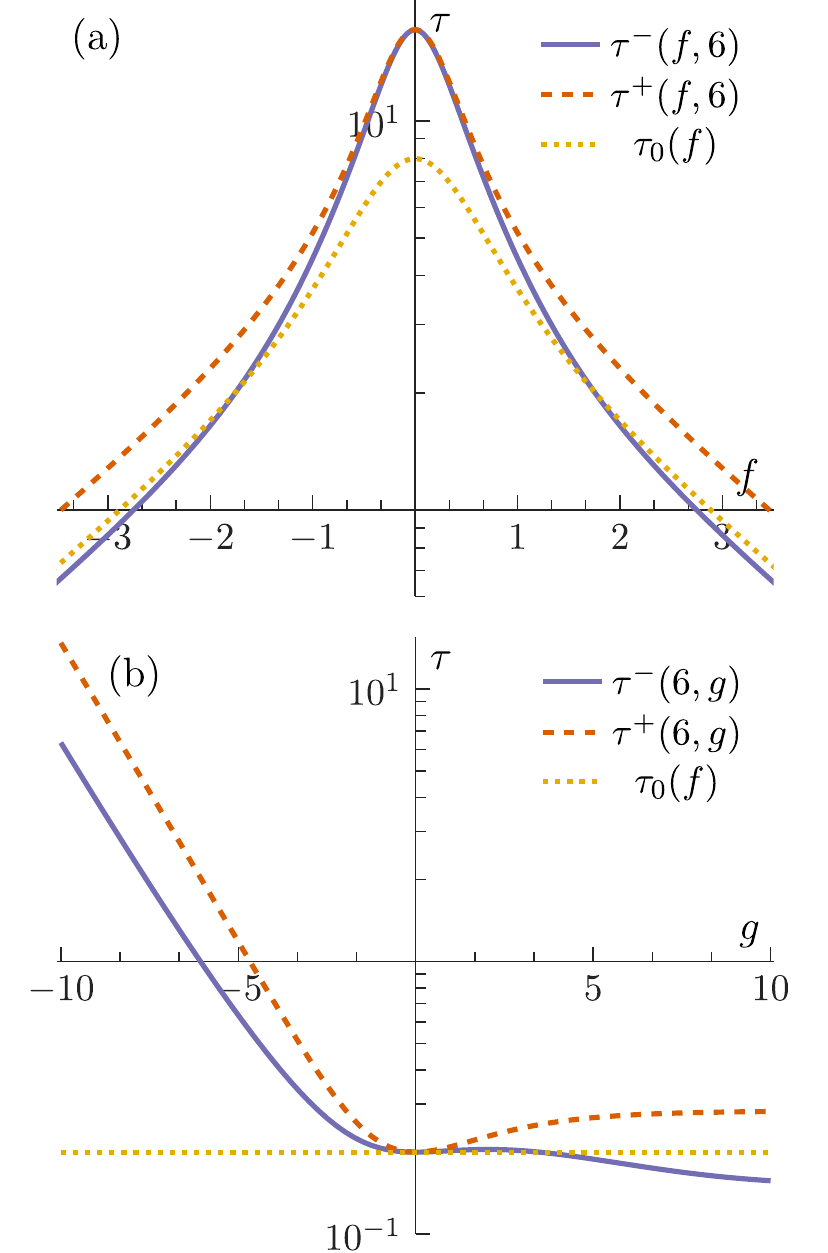} 
\caption{Inequality of cycle completion times in an environment with the same mean particle flow as in the system ($f_{\rm e}=f$). (a) 
Decrease of times with $|f|$ shown for 
repulsive interactions $g=6$. (b) Ordering of times in dependence of the interaction energy $g$ for a bias $f=6$.}
\label{fig:fe=f}
\end{figure}

Figure~\ref{fig:fe=f} shows numerical results for the mean cycle times 
$\tau^+(f,g)$ and $\tau^-(f,g)$ when $f_{\rm e}=f$. For comparison, we also included the mean cycle time 
$\tau_0(f)$ for the non-interacting case ($g=0$), which is the same 
in and against bias direction, $\tau_0(f)=\tau^\pm(f,g=0)$.
Figure~\ref{fig:fe=f}(a) demonstrates the typical dependence on the bias $f$, which we find for any $g$: 
The mean cycle times are maximal at $f=0$ and for large $|f|$
decrease exponentially.  
Figure~\ref{fig:fe=f}(b) demonstrates the ordering of
cycle times for different interactions $g$.

The main findings for $f_{\rm e}=f$ are:

\noindent
\textit{\textbf {(i)} Interaction slows down forward cycles.}\\
For both attractive ($g<0$) and repulsive ($g>0$) interaction (see dashed lines in Fig.~\ref{fig:fe=f}), we have
\begin{equation}\label{eq:interactionslowsdown}
\tau^+(f,g)>\tau_0(f)\,.
\end{equation}
This inequality originates from a lowering of transition rates when the two particles occupy neighboring sites.
In the extreme case $g\to\infty$, the particles would mutually block each other, corresponding to a single-file diffusion, where similar effects have been reported recently for times of single transitions in crowded environments \cite{Ryabov/etal:2019, Shin/etal:2020}. 
For attractive interactions, the particles form a low-energy bound state when encountering each other. To escape from such a bound state requires a longer time on average. This leads to an exponential increase of $\tau^\pm$ for $g\to-\infty$.

\noindent
\textit{\textbf {(ii)} Interaction can accelerate backward cycles.}
The cycle times behave differently for attractive and repulsive interactions.
In particular, for positive  $g$ exceeding a critical value $g_\times$, the inequality 
\begin{equation}\label{eq:acceleration}
\tau^-(f,g)<\tau_0(f)\,,
\end{equation}
holds. This can be understood as follows.
If the system particle wants to complete a backward cycle, the environment particle
is moving in the opposite direction as $f_{\rm e}=f$. Encounters of the two particles can thus hardly be avoided.
For attractive interaction, the formation of low-energy bound states then always leads to $\tau^-(f,g)>\tau_0(f)$. For repulsive interaction, the situation is more subtle as there are two competing effects: First, 
we have the same effect of encounters slowing down the cycle completion. However, 
encounters now more strongly reduce the probability of cycle completion because the environment
particle for $g>0$ has the tendency to block  the motion of the system particle in backward direction. 
This implies that trajectories of the system particle with a small number of encounters, 
which must have a short completion time, become more strongly weighted in the average $\tau^-$. 
This reweighting towards smaller completion times dominates over the slowing down effect for $g>g_\times$ 
and accordingly $\tau^-$ becomes smaller than $\tau_0$.

\noindent
\textit{\textbf {(iii)} Backward cycles are faster than forward ones.}
For $f_{\rm e}=f$, the forward cycle time is never shorter than the backward one,
\begin{equation}\label{eq:splitting}
\tau^+(f,g) \geq \tau^-(f,g)\,.
\end{equation}
The equality is valid for vanishing bias $f=0$ only. Inequality~\eqref{eq:splitting}
holds 
because, in forward direction, the bias helps the system particle to overcome blocking or dragging by the environment particle and the interference by the environment is thus weaker. Accordingly, the average number of encounters for forward cycles is larger than for backward ones. As encounters increase the cycle times (see point (i) above), we obtain the inequality \eqref{eq:splitting}.

\begin{figure}[t] 
\centering 
\includegraphics[scale=1]{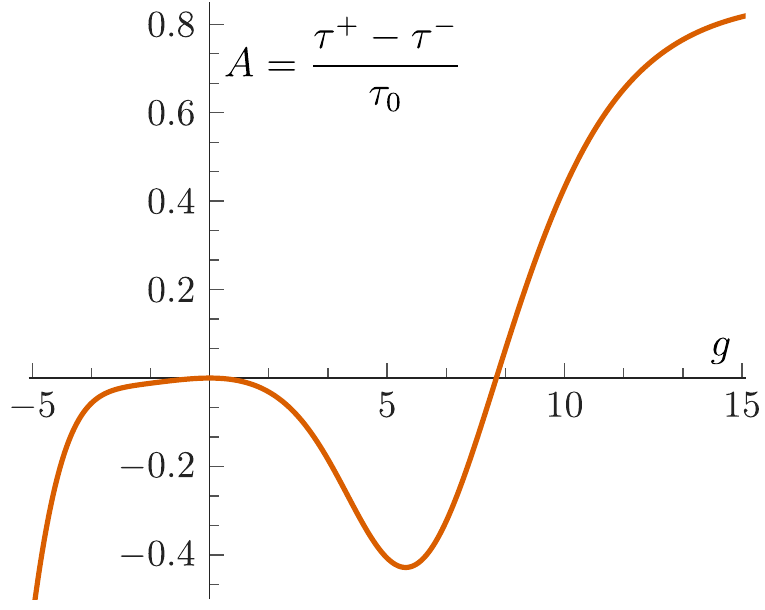} 
\caption{Asymmetry of cycle completion times in the unbiased environment ($f_{\rm e}=0$) in dependence of the interaction energy
$g$ for bias $f=4$.}
\label{fig:asymmetry}
\end{figure}

When the flow in the environment becomes weaker, the difference in the disturbing effect for forward and backward cycling should be reduced 
and thus also the asymmetry between the cycle times, which we quantify by the asymmetry parameter $A=(\tau^+-\tau_-)/\tau_0$. The inequality \eqref{eq:splitting} means $A>0$.

Interestingly enough, without a net flow ($f_{\rm e}=0$) a reversed asymmetry $A<0$ is present for a wide range of parameters $f$ and $g$, where 
\begin{equation}\label{eq:taum<taup}
\tau^+(f,g) < \tau^-(f,g)\,.
\end{equation}
This is exemplified in Fig.~\ref{fig:asymmetry}, which shows that $A$ becomes positive for large positive $g$ only.\\

\begin{figure}[t] 
\centering 
\includegraphics[scale=1]{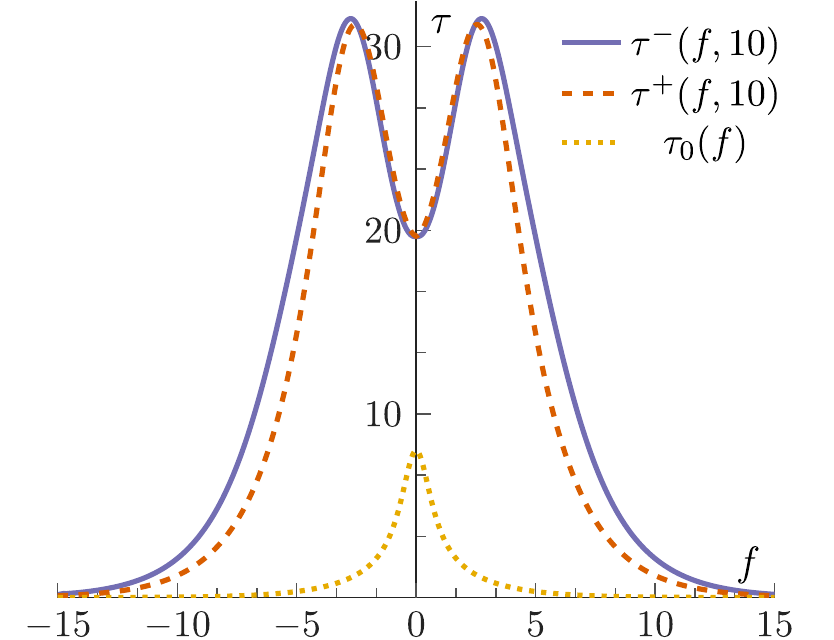} 
\caption{Bimodal shapes of cycle completion times in an environment with the opposite flow compared to the system ($f_{\rm e}=-f$), and for repulsive interaction with $g=10$.}
\label{fig:bimodal}
\end{figure}

What happens when the flow in the environment is opposite to the one in the system? Investigating the case $f_{\rm e}=-f$, we have found a remarkable complex behavior emerging for positive $g$. For small $g$, the times $\tau^{\pm}$ are unimodal functions of $f$ (with single maximum at $f=0$), as in Fig.~\ref{fig:fe=f}.
Upon increasing $g$, a bifurcation occurs at a certain $g_{\rm b}$, where the unimodal function changes into a bimodal one (with two maxima at $f=\pm f_{\rm m}(g)$ and a local minimum at $f=0$). With further increasing $g$, the two maxima become more pronounced. The bimodal shapes are illustrated in Fig.~\ref{fig:bimodal}.\\

Up to now, we have discussed results for a homogeneous situation, where all free energies of the states in the cycle and the energetic barriers between them are equal. Further calculations have shown that all reported features can be found also for
general inhomogeneous cycles in steady states. This holds true for both varying pre-exponential 
factors $a_{ij}$ (energy barriers) and varying state energies $\varepsilon_i$ in Eq.\eqref{eq:kij}

In summary, we have analyzed a basic model for cycle processes interacting with an environment that can have a net particle flow. Our results suggest that mean times $\tau^+$ and $\tau^-$
for completing cycles in forward and backward direction can be rather valuable
quantities to probe couplings and processes in the near environment. 

In particular, a slowing down of a cycle process upon biasing, e.g., by enhancing ATP hydrolysis,
suggests that the net flow in the environment is opposing the preferred cycle direction.
Furthermore, interactions break the symmetry $\tau^+=\tau^-$ of the cycle times and
the sign of the interaction-induced asymmetry ($\tau^+>\tau^-$ or $\tau^+<\tau^-$)
provides information on the type of interaction and the environmental flow. 
An asymmetry $\tau^->\tau^+$ indicates attractive or weak repulsive interaction combined 
with a weak net environmental flow or an equilibrium situation without net flow.  When times for backward cycles in a dense environment
(strong interactions) are shorter than in a dilute system (weak interactions),  this indicates a
repulsive interaction together with a strong net environmental flow in forward direction. 
In this case, $\tau^+$ is larger than $\tau^-$.

\begin{acknowledgement}
Financial support by the Czech Science Foundation (Project No.\ 20-24748J) and the Deutsche Forschungsgemeinschaft (Project No.\ 397157593) is gratefully acknowledged. We sincerely thank the members of the DFG Research Unit FOR 2692 for fruitful discussions.
\end{acknowledgement} 

\bibliography{references_cycle_times} 
\end{document}